\documentstyle[aps,prl,epsfig]{revtex}
\begin{document}

\title{ Analytic Solution of Strongly Coupling Schr\"odinger Equations }
\author{Jinfeng Liao and Pengfei Zhuang\\
        Physics Department, Tsinghua University, Beijing 100084, China }
\maketitle

\begin{abstract}
A recently developed expansion method for analytically solving the
ground states of strongly coupling Schr\"odinger equations by
Friedberg, Lee and Zhao is extended to excited states and applied to the
pedagogically important problems of power-law central forces.
With the extended method, the Hydrogen atom problem is resolved and the
low-lying states of Yukawa potential are approximately obtained.
\end{abstract}

{\bf PACS: 03.65.Ge}

\section{Introduction }

The main task in application of quantum mechanics is to solve
Schr\"odinger equations for various potentials. Unfortunately,
only few of them can be solved exactly. In textbooks various approximation
methods, especially the perturbation theory,
are discussed with great length. While
the perturbation theory, as the most important approximation
method, has exhibited its power in a lot of weakly coupling cases,
it is not applicable to strongly coupling problems which, however,
exist widely and play important roles in the real world. How to
analytically solve the Schr\"oedinger equations in nonperturbative
case remains an open question.

Recently, a new method to discuss strongly
coupling potentials was presented by Friedberg, Lee and
Zhao\cite{flz1}. The goal is to solve the ground state of the
time independent Schr\"oedinger equation with strongly
coupling potential
\begin{equation}
\label{seq}
(\frac{\hat{\vec p}}{2m}+V)\psi=E \psi\ .
\end{equation}
Considering a boundary potential $V(\vec r) \ge 0$ with the
minimum $V(\vec r)=0$ at the origin point $\vec r=0$, one introduces a
dimensionless scalar parameter g which reflects the intensity of
the potential by writing
\begin{equation}
\label{vg}
V(\vec r)=g^{2}v(\vec r)\ .
\end{equation}
For strong potentials we are interested in, $g$ is a large number, the
ground state $\psi$ and the corresponding ground energy $E$ can be expanded
in the inverse number of $g$,
\begin{eqnarray}
\label{exp1}
&&  \psi (\vec r)=e^{-S(\vec r)}\ ,  \nonumber    \\
&&  S(\vec r)=g S_{0} (\vec r)+g^{0} S_{1} (\vec r)+g^{-1} S_{2}
(\vec r)+ \cdot \cdot
 \cdot  \nonumber   \\
&& E=g E_{0}+g^{0} E_{1}+g^{-1} E_{2}+ \cdot \cdot \cdot   \ .
\end{eqnarray}
Substituting the expansion into (\ref{seq}) and then comparing the
coefficients of $g^{-n}$ on the both sides, one obtains a series of
first order differential equations(by contrast, the Schr\"odinger
equation itself is of second order):
\begin{eqnarray}
\label{exp2}
&&(\bigtriangledown S_{0})^{2}=2mv, \nonumber \\
&&\bigtriangledown S_{0} \cdot \bigtriangledown S_{1}=\frac{1}{2}
\bigtriangledown^{2} S_{0}-mE_{0}, \nonumber \\
&&\bigtriangledown S_{0} \cdot \bigtriangledown S_{2}=\frac{1}{2}
(\bigtriangledown^{2} S_{1}-\bigtriangledown S_{1} \cdot
\bigtriangledown S_{1})-mE_{1},  \nonumber \\
&&   \cdot \cdot \cdot
\end{eqnarray}
where $m$ is the particle mass. The first equation of the above
hierarchy contains only $S_{0}$ as unknown term and can be
solved by direct integration with boundary condition. The
second equation contains only $S_{1}$ as unknown term (note that
$S_{0}$ has been solved by the first equation) and can be solved
in the same way as $S_{0}$, and $E_{0}$ can be fixed by
considering the convergence of $S_{1}$ at the origin. With the
same procedure, all $S_{i}$ and $E_{i}$ can be determined step by step. To
illustrate the procedure clearly, one considers the
familiar one dimension harmonic oscillator with $V(x)=g^2 m^3
x^2$. In this case the hierarchy (\ref{exp2}) is reduced to
\begin{eqnarray}
\label{expexp}
&&(\frac{dS_{0}}{dx})^{2}=2m^4 x^2, \nonumber \\
&&\frac{dS_{0}}{dx} \cdot \frac{dS_{1}}{dx}=\frac{1}{2}
\frac{d^2 S_{0}}{dx^2}-mE_{0}, \nonumber \\
&&\frac{dS_{0}}{dx} \cdot \frac{dS_{2}}{dx}=\frac{1}{2} (\frac{d^2
S_{1}}{dx^2}-\frac{dS_{1}}{dx} \cdot
\frac{dS_{1}}{dx})-mE_{1}\ .  \nonumber \\
&&   \cdot \cdot \cdot
\end{eqnarray}
Using the boundary condition $\psi(\pm\infty)=0$ or
$S_0(\pm\infty)=-\infty$ and the convenient renormalization
$\psi(0)=1$ or $S_0(0)=0$, the integration of the first equation
gives $S_0(x)=-\frac{1}{\sqrt 2} m^2 x^2$. With the solved $S_0$, the
second equation is reduced to
$S^{\prime}_{1}=(\frac{1}{2}-\frac{E_0}{\sqrt{2} m})/x$. To avoid
divergence of $S^{\prime}_1$ at $x=0$, the only way is to
choose $E_0=\frac{m}{\sqrt 2} $. With the same procedure, all
$S_{i}$ and $E_{i}$ with $i>0$ can be proven to be zero, and the
final result gives precisely the same solution as that in any
textbook. For details, see \cite{flz1}-\cite{flz3}.

The readers who are familiar with the well-known WKB approximation
may ask the difference between the WKB and the hierarchy (\ref{exp2}).
While the
hierarchy (\ref{exp2}) looks similar to the WKB
hierarchy\cite{merzbacher}, they are very different from each
other. The most explicit difference is that in WKB the energy $E$
appears always in the fist equation of the hierarchy together with
the potential $V$, but here $E_0$ and $v$ may not appear in the
first equation simultaneously, see (\ref{exp2}). In WKB, the
energy is determined by the continuity condition of wave functions
at the so-called "classical turning point", or equivalently by
using the Bohr-Sommerfeld like quantization condition. In the
approach here, the energy eigenvalue is directly determined by the
boundary condition. Physically, in WKB the wave function is
expanded in $\hbar$, it is useful only for semiclassical systems,
while the wave function and energy eigenvalue here are expanded in
the inverse number of the strong coupling constant, the
hierarchy (\ref{exp2}) is like the expansion in the inverse number
of large $N$ often used in nonperturpative treatment of quantum
field systems, see, for instance, \cite{witten}. Also the WKB wave
function $e^{{i\over \hbar}\alpha(x)}$ is an oscillating form
which corresponds to real particles, so that more applicable to
semiclassical case, but the wave function in (\ref{exp2}) is
a damping form and more suitable for dealing with bound states.

Although the new method is powerful in solving strongly coupling
Schr\"odinger equations, the step of writing the wave function
$\psi (\vec r)$ in the form $\psi (\vec )=e^{S(\vec r)}$ in (\ref{exp1}) is a
strong constraint in the application of the method to more
kinds of potentials. For example, the linear potential $V=gr$
cannot be solved through expansion (\ref{exp1}) due to the
divergence encountered in dealing with (\ref{exp2}). From the analysis in
Section 2 we will see
that the expansion with the form of (\ref{exp1}) is only suitable
for harmonic-oscillator-like potentials which can be
expanded quadratically around the minimum. On the other hand, the
expansion (\ref{exp1}) is valid mainly for ground states, an excited state
can not be expressed as a pure exponential function.

To make the expansion method more applicable, one should construct carefully
the expansion form of $\psi (\vec r)$ through
physical analysis of different potentials and of different states.
In this paper, we first investigate
the coupling constant dependence of bound states for power-law
central forces in order to construct proper expansions of $\psi$ and $E$
in inverse coupling
constant in Sections 2 and 3, and then apply the extended expansion method to
Coulomb potential and resolve the excited states in Section 4.
Finally we discuss Yukawa potential and obtain the low-lying
states approximately in Section 5. The Conclusions are given in
Section 6. We have chosen $\hbar=c=1$ throughout the paper.

\section{ Coupling constant dependence of bound states solutions }

As mentioned above, the dependence of bound states solutions on
coupling constant is crucial since the method is based on the
coupling constant expansion. Qualitatively speaking, when coupling
constant $g$ in potentials like (\ref{seq}) increases, the bound state
energy drops down and the space extension of the corresponding
wave function becomes narrower, that is, the system is more
bounded. In the following we concentrate our discussion on
power-law central forces $V(r) \propto g^k r^n$, the reason is
twofold: first, with those potentials the coupling constant
dependence can be factorized, as we will see below; second, any
boundary potential closely around its minimum point can be
approximated well by power-law curve with certain value of $n$.

\subsection{ Coupling constant dependence of energy }

Let's consider a power-law central potential
\begin{eqnarray}
\label{power}
&& \hat{H}=\frac{\hat p^2}{2m} + V \ ,\nonumber\\
&& V(r)=\pm g^k m (mr)^n\ .
\end{eqnarray}
In order to keep boundary condition we should choose the sign $+$ for $n>0$
and $-$ for $n<0$.
Here the coupling constant $g$ is guaranteed to be
dimensionless.  For instance, the harmonic oscillator $V=\frac{1}{2} m
\omega^2 r^2 $ can be written as $V=g^k m^3 r^2$ with
$g^k=\frac{1}{2} (\frac{\omega}{m})^2$.

Considering $g$ as a parameter in the Hamiltonian, we apply Hellmann-Feynman
Theorem\cite{merzbacher} for any arbitrary energy level E and the
corresponding state $|\psi \rangle$
\begin{equation}
\label{hellmann}
\frac{\partial E}{\partial g}=\langle \psi | \frac{\partial
\hat{H}}{\partial g} | \psi \rangle
\end{equation}
to (\ref{power}) and obtain
\begin{equation}
\label{dedg}
\frac{\partial E}{\partial g}=\frac{k}{g} \langle \psi | V | \psi
\rangle .
\end{equation}
From Virial Theorem\cite{merzbacher},
\begin{equation}
\label{virial}
\langle \psi | V | \psi \rangle =\frac{2}{n+2} \langle \psi |
\hat{H} | \psi \rangle = \frac{2}{n+2} E ,
\end{equation}
we get
\begin{equation}
\label{dedg2}
\frac{\partial E}{\partial g}=\frac{2k}{n+2} \frac{1}{g} E ,
\end{equation}
and its solution
\begin{equation}
\label{fac}
E(g)=g^{\frac{2k}{n+2}} \epsilon ,
\end{equation}
where $\epsilon$ depends on $r$ only, the coupling constant
dependence is factorized. A useful consequence of this
factorization is that for a given potential $V(r)$ one can
determine firstly $\epsilon$ in some ideal case, for instance, the
limit $g \gg 1$ or $g \ll 1$, and then obtain the energy $E$ of
the real system by multiplying $\epsilon$ by the $g$-factor
$g^{2k\over n+2}$.

When a potential is not in the power-law form, the factorization
(\ref{fac}) fails. However, in strongly coupling cases
the low-lying states are restricted in a small region
around the origin $r=0$, and the potential can be expressed in
the power-law form
\begin{equation}
\label{pl}
\lim_{r \to 0} V(r) \propto \pm g^k m (mr)^n
\end{equation}
in the neighborhood of the origin, the factorization is still
approximately valid, and the leading term $E_0$ in the expansion
(\ref{exp1}) is given by (\ref{fac}). This conclusion will be used
to solve the Yukawa potential
\begin{equation}
\label{yukawa}
V(r)=-g^2 e^{-\alpha r}/r
\end{equation}
in Section 5.

\subsection{ Scale transformation }

For central forces, the bound state wave functions can be written
as\cite{merzbacher}
\begin{equation}
\label{phi} \psi(\vec{r})= R(r) Y_{LM}(\theta,\phi) ,
\end{equation}
where $Y_{LM}(\theta,\phi)$ is the spherical harmonic function
carrying quantum numbers $L$ and $M$, and $R(r)$ which is related
to the coupling constant satisfies the radial equation
\begin{equation}
\label{radial}
\frac{d^2 R(r,g)}{dr^2}+\frac{2}{r}
\frac{dR(r,g)}{dr}+2m(E-V)R(r,g)-\frac{L(L+1)}{r^2}R(r,g)=0 .
\end{equation}
Considering a scale transformation
\begin{eqnarray}
\label{sca1}
   && r \to r_s = ar ,   \nonumber \\
   && g \to g_s = bg ,
\end{eqnarray}
with a and b being two positive and real constants, and using the
factorization (\ref{fac}) the radial equation (\ref{radial}) is
transformed into
\begin{equation}
\label{radial2}
\frac{1}{a^2} \frac{d^2 R(r_s,g_s)}{dr^2}+ \frac{1}{a^2}
\frac{2}{r} \frac{dR(r_s,g_s)}{dr} +
2m(b^{\frac{2k}{n+2}}E-b^k a^n V ) R(r_s,g_s) - \frac{1}{a^2}
\frac{L(L+1)}{r_s^2} R(r_s,g_s)=0 .
\end{equation}
If the constants $a$ and $b$ are restricted by
\begin{equation}
\label{sca2}
b^{\frac{2k}{n+2}} = b^k a^n = a^{-2} ,
\end{equation}
$R(r_s,g_s)$ and $R(r,g)$ satisfy the same radial equation and the
same boundary conditions at the origin and at infinity( the
transformation does not change boundary conditions ), that is
\begin{equation}
\label{sca3}
R(r_s,g_s) = c R(r,g) ,
\end{equation}
where c is a $g$ and $r$ independent constant and can be removed
by normalization.

The invariance of the equation of motion under the scale
transformation (\ref{sca1}) and (\ref{sca2}) originates from the
symmetry of the Hamiltonian ({\ref{power}),
\begin{equation}
\label{hs}
\hat{H}  \to \hat{H_s} = \frac{1}{a^2} \hat{H} .
\end{equation}
Under the transformation all eigenvalues of the Hamiltonian are
rescaled by a constant factor $1/a^2$ and all eigenfunctions are
unchanged.

The scale transformation property of the radial function implies
that we can replace the original two variables $r$ and $g$ by one
dimensionless scale variable
\begin{equation}
\label{s} s = g^k (mr)^{n+2}
\end{equation}
which is invariant under the scale transformation (\ref{sca1}) and (\ref{sca2}).
With properly selected normalization scheme, the radial functions $R(s_1)$
and $R(s_2)$ with two different coupling constants $g_1$ and $g_2$ are related by
\begin{eqnarray}
\label{rs} R(g_1^k (mr)^{n+2}) &=& R \left( g_2^k \left({g_1\over
g_2}\right)^k (mr)^{n+2} \right) \nonumber \\
&=& R \left( g_2^k\left(\left({g_1\over g_2}\right)^{k\over
n+2}mr\right)^{n+2}\right) .
\end{eqnarray}
Therefore, the wave function at point $r$ with coupling constant
$g_1$ equals the wave function at point $(g_1/g_2)^{k/n+2} r$ with
$g_2$. The wave function is contracted in a smaller region when
the coupling becomes stronger while extended to a larger region
when the coupling becomes weaker. With the relation (\ref{rs}),
the behavior of a weakly coupling system at the place far from the
origin is equivalent to that of a strongly coupling system near
the origin.

The conclusion of this section can be summarized as
\begin{eqnarray}
\label{sum}
&& V= \pm g^k m (mr)^n \quad , \nonumber \\
&& E= g^{\frac{2k}{n+2}} \epsilon \quad , \nonumber \\
&& R(r,g)=R(g^k (mr)^{n+2}) .
\end{eqnarray}
For the potentials with power-law behavior (\ref{pl}) around the
origin, the leading terms of the energy and wave function are also
given by (\ref{sum}). This conclusion tells us how to write down
the expansion in $g$ for a power-law or a power-law-like central
potential. For example, the expansion (\ref{exp1}) is only
suitable for harmonic oscillator and the similar potentials.

\section{ Extended expansion method }

The key point of the method by Friedberg, Lee and Zhao is to
expand the energy and wave function of a strongly coupling system
in the inverse number of the coupling constant $g$. As mentioned
above, the original expansion (\ref{exp1}) is 1) only suitable for
the potentials with harmonic oscillator behavior around the
minimum, and 2) only valid for ground states. To solve (\ref{seq})
by the expansion method, we should proceed three steps: First,
write a proper g-power expansion for energy and wave function
according to the scaling property discussed in previous section;
Second, substitute the expansion back into (\ref{seq}) and obtain
a hierarchy of first-order differential equations; Last, integrate
the hierarchy one by one with the help of the boundary and
convergence conditions, as described in the example of one
dimension harmonic oscillator in Section 1.

As a pedagogical goal, let's extend the method to solve three
dimensional Coulomb and harmonic oscillator problems in quantum
mechanics.

In \cite{flz1} and \cite{flz2}, the expansion and solution of the
ground state are given by
\begin{eqnarray}
\label{cou1}
V &=&-g^2 \frac{1}{r} \nonumber \\
E &=&g^4 E_0 + g^2 E_1 + g^0 E_2 + \cdot \cdot \cdot  \nonumber \\
  &=&-{1 \over 2}g^4 m\ \nonumber\\
S(r) &=& g^2 S_0 (r) + g^0 S_1 (r) + g^{-2} S_2 (r) + \cdot \cdot \cdot\nonumber\\
     &=& g^2 m r\
\end{eqnarray}
for Coulomb potential, and
\begin{eqnarray}
\label{har1}
V &=&\frac{1}{2} g^2 m^3 r^2   \nonumber\\
E &=&g E_0 + g^0 E_1 + g^{-1} E_2 + \cdot \cdot \cdot  \nonumber \\
  &=&{3\over 2}g m\ \nonumber\\
S(r) &=& g S_0 (r) + g^0 S_1 (r) + g^{-1} S_2 (r) + \cdot \cdot \cdot\nonumber\\
     &=& {1 \over 2} g  m^2 r^2\
\end{eqnarray}
for harmonic oscillator. It is clear that the expansions for the
two potentials are in accord with the scaling law shown in
(\ref{sum}).

Since any excited state is not so bounded like the ground state,
we assume that the excited states differ from the ground state
$e^{-S(r)}$ by a smooth function $P(r)$,
\begin{equation}
\label{exc1} R(r)=P(r) e^{-S(r)}\ .
\end{equation}
When $P(r) =1$ we go back to the ground state. From the spirit of
the expansion method we express $P(r)$ and $S(r)$ as finite
polynomials of $1/g$. With the help of the scaling law
(\ref{sum}), they can be further rewritten as polynomials of a
dimensionless scaling variable $g^2 m r$ for Coulomb potential
\begin{eqnarray}
\label{cou2}
&& S(r)= \sum_{j=1}^{-\alpha} b_{1-j}(g^2 m r)^j \ , \nonumber \\
&& P(r)= \sum_{i=\beta}^{-\gamma}a_i (g^2 m r)^i \ , \nonumber \\
&& E=g^4 \epsilon\ ,
\end{eqnarray}
or $g m^2 r^2$ for harmonic oscillator
\begin{eqnarray}
\label{har2}
&& S(r)= \sum_{j=1}^{-\alpha}  b_{1-j}(g m^2 r^2)^{j}\ ,  \nonumber \\
&& P(r)= \sum_{i=\beta}^{-\gamma} a_i (g m^2 r^2)^{i}\ ,  \nonumber \\
&& E=g \epsilon\ .
\end{eqnarray}
Here $\alpha,\beta,\gamma$ are nonnegative integers, and $a_i,
b_{1-j}$ are $g$ and $r$ independent constants. Note that the
expressions (\ref{cou2}) and (\ref{har2}) obtained from the
scaling analysis guarantee a wider spread of excited states than
the ground state. The expansions here are also in accordance with
the discussion with algebra method\cite{ajp}.

We now substitute the expansions into the corresponding radial
equations and discuss the convergence in the limit $r \to 0$. It
is easy to prove that when $\alpha > 0$ the most divergent term in
the radial equation is $ \alpha^2 g^{-4\alpha} b_{\alpha+1}^2
r^{-(\gamma+2\alpha+2)} $ for Coulomb potential and $ 4 \alpha^2
g^{-2\alpha} b_{\alpha+1}^2 r^{-2(\gamma+2\alpha+1)} $ for
harmonic oscillator. Since there is only one most divergent term,
it is impossible to cancel it by other terms. The only way is to
take $\alpha = 0$. In this case, the expansion of $S(r)$ in
(\ref{cou2}) or (\ref{har2}) is reduced to two terms, and the
constant term can be absorbed by the normalization of the wave
function. Moreover, the terms with negative powers ($\gamma > 0$)
in the expansion of $P(r)$ must vanish also in order to keep the
whole radial wave function $R(r)=P(r) e^{-S(r)}$ from divergence
at the origin $r=0$. Finally we obtain for any bound state the
extended expansion
\begin{eqnarray}
\label{cou4}
&& S(r)= b_0 g^2 m r\ ,\nonumber \\
&& P_N (r) = \sum_{i=0}^{N-1} a_i (g^2 m r)^i
\end{eqnarray}
for Coulomb potential, and
\begin{eqnarray}
\label{har4}
&& S(r) = b_0 g m^2 r^2 \nonumber \\
&& P_N (r) = \sum_{i=0}^{N-1} a_k ( g m^2 r^2 )^i
\end{eqnarray}
for harmonic oscillator with $N\ge 1$.

\section{ Solutions for Coulomb potential}

We now turn to the second step, namely substitute the convergent
expansion (\ref{cou4}) into the radial equation for Coulomb
potential, and get a series of equations in different orders of
$g$:
\begin{eqnarray}
\label{cou5}
 g^{2N+2} :&& \quad b_0^2 m + 2 \epsilon = 0 \nonumber \\
 g^{2N}   :&& \quad 1 - b_0 N = 0  \nonumber \\
 g^{2k} (0<k<N) : && \left(k(k+1)-L(L+1)\right) a_k + \frac{2}{N} (N-k) a_{k-1} = 0  \nonumber \\
 g^0  : && \quad L(L+1) a_0 = 0
\end{eqnarray}

From the first two equations we derive immediately the energy
level $\epsilon_N = - \frac{m}{2 N^2}$ and the coefficient $b_0 =
\frac{1}{N} $ in $S$.

To obtain the coefficients $a_i$ in the polynomial $P$, we need to
discuss the relation between the two quantum numbers $N(\ge 1)$
and $L(\ge 0)$, they come from the polynomial $P_N (r)$ and the
spherical harmonic function $Y_{LM}(\theta,\phi) $. For $N=1$, the
requirement $P_1 (r) = a_0 \ne 0$ leads to $L=0$ from the last
equation of (\ref{cou5}), the general expression (\ref{cou4}) is
then reduced to the ground state solution (\ref{cou1}).

For $N > 1$, we discuss three cases for the quantum number $L$
separately: 1) $L=0$, the third equation of (\ref{cou5})
determines the recursion relation between the coefficients $a_k$
\begin{eqnarray}
\label{swave}
&& R_{N0}= \left( \sum_{k=0}^{N-1} a_k (g^2 mr)^k \right) e^{-g^2 mr / N }  \nonumber \\
&& a_k = - \frac{2(N-k)}{N k (k+1)} a_{k-1} ,\ \ \ \ \ \ \  0 < k < N ,
\end{eqnarray}
the only unknown coefficient $a_0 \ne 0$ is determined by the normalization.
2) $0<L<N$, the last two equations of (\ref{cou5}) lead to the coefficients $a_0
= a_1 = a_2 = \cdot \cdot \cdot =a_{L-1}=0 $, and then the radial solutions
are written as
\begin{eqnarray}
\label{ng0}
&& R_{NL} = \left( \sum_{k=L}^{N-1} a_k (g^2 mr )^k \right) e^{-g^2 mr / N}  \nonumber \\
&& a_k = \frac{2}{N} \frac{N-k}{L(L+1)-k(k+1)} a_{k-1},
\quad L<k<N
\end{eqnarray}
$a_L \ne 0$ is determined by the normalization too. 3) $L \ge N$,
the last two equations of (\ref{cou5}) require $a_0 = a_1 = a_2 =
\cdot \cdot \cdot =a_{N-1}=0$, and there is no nonzero solution in
this case.

We summarize the solution of the bound states for Coulomb
potential:
\begin{eqnarray}
\label{cou6}
&& E_N = - \frac{1}{2N^2} g^4 m \nonumber \\
&& \psi_{NLM}(r,\theta,\phi) = \left( \sum_{k=L}^{N-1} a_k (g^2 mr
)^k \right) e^{-g^2 mr / N  }
   Y_{LM} (\theta, \phi)  \nonumber \\
&& a_k = \frac{2}{N} \frac{N-k}{L(L+1)-k(k+1)} a_{k-1}\ ,
\nonumber\\
&& N>0, \quad L=0,1,2, \cdot \cdot \cdot , N-1
\end{eqnarray}
with $a_L \ne 0$ determined by the normalization of the wave
function. (\ref{cou6}) is exactly the same as what obtained by
solving the second-order Schr\"odinger equation directly in normal
textbooks\cite{merzbacher}, but here by using the expansion in
coupling constant and with the help of the scaling law the
complicated second-order differential equations are replaced by
simple algebra equations.

In a similar way one can also derive all the bound states for harmonic
oscillator by substituting the convergent
expansion (\ref{har4}) into the radial equation.

\section{Yukawa potential}

The potential (\ref{yukawa}) was first introduced into physics in
1930s by Yukawa in the study of strong interaction between
nucleons through meson exchange\cite{yukawa1}. It is also known as
Debye-H\"uckel potential in plasma physics and Thomas-Fermi
potential in solid-state physics\cite{yukawa2}. While the Yukawa
potential is important in physics, the corresponding Schr\"odinger
equation can not be solved analytically and exactly. A lot of
publications have been contributed to the problem using different
approximations (see \cite{wkb2} - \cite{yukawa5} and references
therein). Here we use the extended expansion method to obtain its
ground state and low-lying excited states. Comparison with strict
but numerical calculation will be made.

We first consider the proper $1/g$ expansion for Yukawa potential.
Since Yukawa potential is not a power-law central force, the $g$
dependence of its bound states can not be simply factorized like
(\ref{sum}). In strong coupling case with $g >> 1$, however, the
low-lying wave functions are mainly distributed in the vicinity of
the origin. Since in the limit of $r \to 0$, Yukawa potential
approaches to Coulomb potential, we can thus write the expansion
of the low-lying states of Yukawa potential by properly modifying
the expansion of Coulomb potential. The inverse of the parameter
$\alpha$ in the Yukawa potential (\ref{yukawa}) represents the
mean potential range. To make our calculation based on Coulomb
potential( whose potential range approaches infinity) more
effective, we should require the mean potential range to be much
larger than the "Bohr radius" of Coulomb potential, that is,
$\frac{1}{\alpha}
>> \frac{1}{g^2 m}$, or equivalently $\frac{\alpha}{g^2 m} << 1 $.
We will show below how the dimensionless number $\frac{\alpha}{g^2
m}$ controls the expanding series.

We now deal with the ground state. In the light of (\ref{cou1})
for Coulomb potential, we expand the energy and wave function as
\begin{eqnarray}
\label{yukawa1}
&& E=g^4 E_0 + g^2 E_1 + g^0 E_2 + \cdot \cdot \cdot  \nonumber \\
&& R = e^{-S(r)}  \nonumber \\
&& S(r)= g^2 S_0 (r) + g^0 S_1 (r) + g^{-2} S_2 (r) + \cdot \cdot
\cdot
\end{eqnarray}
Note that the expansions of $E$ and $S$ are finite for strict
Coulomb potential but infinite for Coulomb-like potentials.

By Substituting the expansions (\ref{yukawa1}) into the
corresponding stationary Schr\"odinger equation, we obtain an
infinity hierarchy of first-order differential equations
(\ref{expexp}). With similar procedure in dealing (\ref{expexp}),
we can solve them one by one to any order we want. The energy
obtained to ${\cal O}(g^{-4})$ and the wave function to ${\cal
O}(g^{-2})$ read:
\begin{eqnarray}
\label{yukawa2} && E= -\frac{1}{2} g^4 m + g^2 \alpha -
\frac{3\alpha^2}{4m} +
g^{-2} \frac{\alpha^3}{2 m^2}   +  {\cal O} (g^{-4}) = g^4 m [- \frac{1}{2} +
\frac{\alpha}{g^2 m}-\frac{3}{4} (\frac{\alpha}{g^2 m})^2+ \frac{1}{2}
(\frac{\alpha}{g^2 m})^3+{\cal O} ((\frac{\alpha}{g^2 m})^4)]\ , \nonumber \\
&& S(r) = g^2 mr + \int_{0}^{r} dr' [\frac{1}{r}(1-e^{-\alpha
r'})-\alpha ] + {\cal O}({g^{-2}}) = g^2 m [r + \frac{\alpha}{g^2
m } \int_{0}^{r} dr' (\frac{1-e^{-\alpha r'}}{\alpha r'}-1) +
{\cal O} ((\frac{\alpha}{g^2 m})^2) ] .
\end{eqnarray}
We see clearly that the dimensionless parameter $\alpha/g^2 m$ controls the degree of
convergence of the expansion. When $\alpha$ vanishes, Yukawa potential is reduced to Coulomb
potential, and therefore, the above solution is reduced to the strict solution (\ref{cou1}).

We now consider the lowest excited state. By modifying the expansion (\ref{cou4})
with $N=2$ for Coulomb
potential, we write the following expansion
\begin{eqnarray}
\label{yukawa3}
&& E=g^4 E_0 + g^2 E_1 + g^0 E_2 + \cdot \cdot \cdot  \nonumber \\
&& \psi = P(r) e^{-S(r)} Y_{LM} (\theta,\phi)  \nonumber \\
&& P(r) = g^2 b_1(r) + g^0 b_0(r) \nonumber \\
&& S(r)= g^2 S_0 (r) + g^0 S_1 (r) + g^{-2} S_2 (r) + \cdot \cdot\
. \cdot
\end{eqnarray}
Unlike the expansion for strict Coulomb potential, here $E$ and
$S(r)$ have infinite terms, and the $r$-dependence of $b_0$ and
$b_1$ in $P(r)$ is unknown and only in the limit case we have:
\begin{eqnarray}
\label{limit}
&&\lim_{r \to 0} b_1 (r) = \lim_{\alpha \to 0} b_1
(r) \propto mr \ ,\nonumber \\
&&\lim_{r \to 0} b_0 (r) = \lim_{\alpha \to 0} b_0 (r) \propto 1\
.
\end{eqnarray}

Substituting the expansions (\ref{yukawa3}) into the radial
equation with $N=2$, we get again a series of first-order
differential equations. The first two which are respectively
proportional to $g^6$ and $g^4$ read
\begin{eqnarray}
\label{yukawa4}
&& \frac{dS_0}{dr} = \sqrt{-2m E_0} \ ,\nonumber \\
&& \frac{dS_0}{dr} \frac{dS_1}{dr} =
\frac{1}{b_1}  \frac{d b_1}{dr} \frac{dS_0}{dr} +
\frac{1}{r} ( \frac{dS_0}{dr} -m e^{-\alpha r} ) - m E_1  .
\end{eqnarray}
By using (\ref{limit}) the condition to keep convergence of
$dS_0/dr$ and $dS_1/dr$ at $r=0$ result in
\begin{eqnarray}
\label{yukawa5}
&& E_0 = - \frac{m}{8}\ ,   \nonumber \\
&& \frac{dS_0}{dr} = \frac{m}{2} \ ,\nonumber \\
&& \frac{dS_1}{dr} = \frac{1-2e^{-\alpha r}}{r} + \frac{1}{b_1}
\frac{d b_1}{dr} - 2 E_1\ .
\end{eqnarray}
In a similar way we can solve the equations proportional to $g^2$
and $g^0$ to determine $E_1$ and $S_1$
\begin{eqnarray}
\label{yukawa6}
&& E_1 = \frac{6 + L (L+1)}{8} \alpha \ ,\nonumber \\
&& b_0 = 1 - L  \ ,\nonumber \\
&& b_1 = -\frac{m}{2\alpha} (1 -  e^{-\alpha r} )\ .
\end{eqnarray}
Put the constituents obtained above together we finally get approximately the analytical
solution of the lowest excited state for Yukawa potential as follows
\begin{eqnarray}
\label{yukawa7} && E = - \frac{m}{8} g^4 + \frac{6 + L (L+1)}{8 }
g^2 \alpha  +  {\cal O}(g^{0}) = -\frac{g^4 m}{8}
[1-\frac{\alpha}{g^2 m} (L^2+L+6) + {\cal O} ((\frac{\alpha}{g^2
m})^2)  ]\ ,
\nonumber \\
&& S(r) = g^2 m [\frac{r}{2} + \frac{\alpha}{g^2 m} \int_{0}^{r}
dr' (\frac{1-2 e^{-\alpha r'}}{\alpha r'}+\frac{e^{-\alpha
r'}}{1-e^{-\alpha r'}}-\frac{L^2+L+6}{4} )+ {\cal O}
((\frac{\alpha}{g^2 m})^2)  ] \ , \nonumber\\
&& P(r) = a_0 \left((1-L)-\frac{1}{2}\frac{g^2 m}{\alpha}
(1-e^{-\alpha r})\right)\ ,\nonumber\\ && N=2, \quad L=0,1 \ ,
\end{eqnarray}
the only constant $a_0$ is determined by normalization. Again, the
dimensionless parameter $\alpha/g^2 m$ dominates the convergence.
The analytic result $E$ for ground and lowest excited states is
compared with the strict but numerical result $E_N$ and with the
Coulomb result $E_C$ in Tabel (\ref{tab}) for different values of
the dimensionless parameter $\alpha/g^2 m$. All the energy values
have been scaled by $g^4 m$. Since the ground state energy is
calculated to the fourth order, it always agrees well with the
strict solution. The Coulomb energy is just the leading order of
the Yukawa energy, its deviation from the strict result is very
strong, especially for the excited states. then relatively large.
When the parameter $\alpha/g^2 m$ decreases, the analytic result
to the second order for the excited states becomes more and more
close to the strict solution.

We also compared our energy eigenvalues with the often cited
numerical calculations in \cite{yukawa5}. The deviation of our
analytical result from the high accurate numerical result is less
than $2.5 \%$ for $1s$ state with $\frac{\alpha}{g^2 m}<1/3$, $5
\%$ for $2s$ state with $\frac{\alpha}{g^2 m}<1/30$, and $5 \%$
for $2p$ state with $\frac{\alpha}{g^2 m}<1/20$. The deviation
decreases when the dimensionless parameter $\alpha/g^2 m$ becomes
smaller.

The other excited states ( $N=3,4,\cdot \cdot \cdot$ ) can be
obtained with the similar method.

\begin{figure}[ht] \hspace{+0cm}
\centerline{\epsfxsize=16cm\epsffile{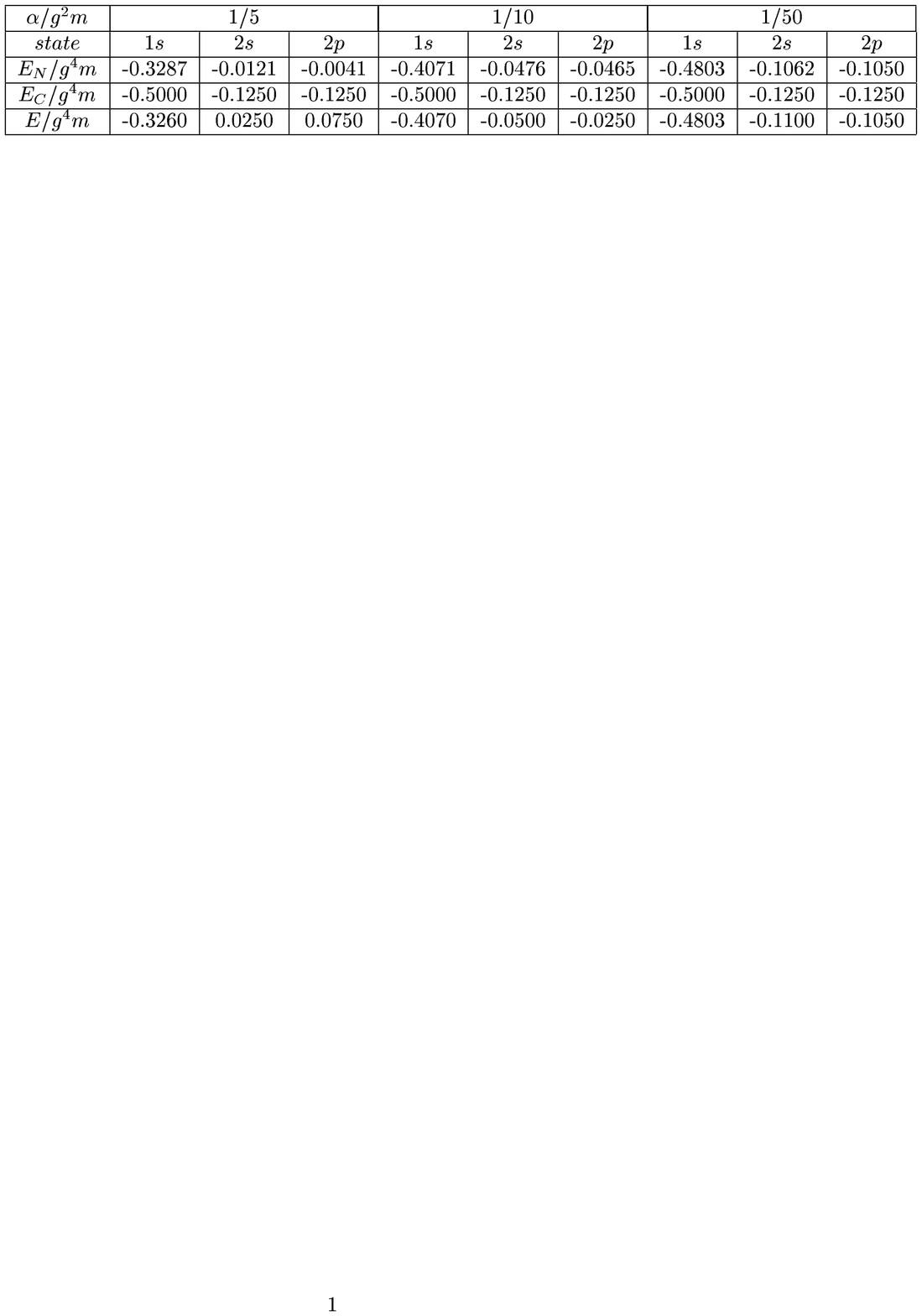}} \caption{\it
Comparison of our analytic result $E$ of Yukawa potential to the
fourth order for the ground state and to the second order for the
lowest excited states with the strict but numerical result $E_N$
and with the Coulomb result $E_C$ for three values of the
dimensionless parameter $\alpha/g^2 m$. } \label{tab}
\end{figure}

The solutions (\ref{yukawa2}) and (\ref{yukawa7}) have the following features: \\
1) The leading orders are just the solutions of Coulomb potential,
and therefore are invariant under the scale transformation
(\ref{sca1}). The high order corrections break the scale
invariance explicitly. When $\alpha \to 0$ or $r \to 0$, the high
order corrections approach to zero
and the scale invariance is restored.\\
2) Different from the accidental degeneracy for Coulomb potential, the energy level for
Yukawa potential depends on the quantum number $L$.\\
3) The expanding series are controlled by the combined
dimensionless parameter $\frac{\alpha}{g^2 m}$. For sufficiently
small values, namely for very strong interaction, the expansions
(\ref{yukawa2}) and (\ref{yukawa7}) converge very fast, one can
consider the lower-order contributions only.

\section{Summary}

The recently developed expansion method\cite{flz1} - \cite{flz3}
provides an alternative way to solve analytically Schr\"oedinger
equations with strong coupling. The key point of the method is to
construct proper expansions for different potentials and different
quantum states. Through investigating the scale transformation
invariance of the bound states for power-law central forces, we
have extended the expansion method from ground state to any
excited state for power-law central forces and to low-lying states
for power-law-like central forces. With the extended expansion
method, we obtained analytically the strict solutions of all bound
states for Coulomb potential and the approximate solutions of
low-lying states for Yukawa potential. Further application of the
method to other physical potentials is of interest and value.

\section*{Acknowledgements}

We are grateful to H. Zhai and W.Q.Zhao for valuable discussions.
The work was supported in part by the NSFC and the project 973 of
China.

\end{document}